\documentclass[conference]{IEEEtran}
\IEEEoverridecommandlockouts
\usepackage{cite}
\usepackage{amsmath,amssymb,amsfonts}
\usepackage{algorithmic}
\usepackage{textcomp}
\def\BibTeX{{\rm B\kern-.05em{\sc i\kern-.025em b}\kern-.08em
    T\kern-.1667em\lower.7ex\hbox{E}\kern-.125emX}}
\usepackage[table,xcdraw]{xcolor}
\usepackage{multirow}
\usepackage[super]{nth}
\usepackage{graphicx,adjustbox}
\usepackage{caption}
\usepackage[labelformat=simple]{subcaption}
\usepackage{comment}
\usepackage{setspace}
\usepackage{textcomp}
\usepackage{xspace}
\usepackage{siunitx}
\usepackage{epsfig}
\usepackage{epstopdf}
\usepackage{soul}
\usepackage{url}
\usepackage{tablefootnote}
\usepackage[linesnumbered,ruled]{algorithm2e}
\usepackage{booktabs}
\usepackage{hyperref}
\usepackage[normalem]{ulem}
\usepackage{footnote}
\usepackage[misc,geometry]{ifsym} 
\usepackage{changepage}
\usepackage{mathtools}

\begin{document}

\title{Towards Energy Efficient LPWANs through Learning-based Multi-hop Routing
\thanks{This work has been partially supported by the Spanish Ministry of Economy and Competitiveness under the Maria de Maeztu Units of Excellence (MDM-2015-0502), by the Spanish Government through the project TEC2016-79510-P, and by the Catalan Government through the projects 2017SGR 1188 and 2017SGR 1739. The work done by S. Barrachina-Mu\~noz is supported by a FI grant from the Generalitat de Catalunya.}
}

\author{\IEEEauthorblockN{1\textsuperscript{st} Sergio Barrachina-Mu\~noz}
\IEEEauthorblockA{\textit{Wireless Networking} \\
\textit{Universitat Pompeu Fabra}\\
Barcelona, Spain \\
sergio.barrachina@upf.edu}
\and
\IEEEauthorblockN{2\textsuperscript{nd} Toni Adame}
\IEEEauthorblockA{\textit{Network Tech. and Strategies} \\
\textit{Universitat Pompeu Fabra}\\
Barcelona, Spain \\
toni.adame@upf.edu}
\and
\IEEEauthorblockN{3\textsuperscript{rd} Albert Bel}
\IEEEauthorblockA{\textit{Network Tech. and Strategies} \\
\textit{Universitat Pompeu Fabra}\\
Barcelona, Spain \\
albert.bel@upf.edu}
\and
\IEEEauthorblockN{4\textsuperscript{th} Boris Bellalta}
\IEEEauthorblockA{\textit{Wireless Networking} \\
\textit{Universitat Pompeu Fabra}\\
Barcelona, Spain \\
boris.bellalta@upf.edu}
}

\maketitle

\begin{abstract}
	
	Low-power wide area networks (LPWANs) have been identified as one of the top emerging wireless technologies due to their autonomy and wide range of applications. 
	Yet, the limited energy resources of battery-powered sensor nodes is a top constraint, especially in single-hop topologies, where nodes located far from the base station must conduct uplink (UL) communications in high power levels.
	On this point, multi-hop routings in the UL are starting to gain attention due to their capability of reducing energy consumption by enabling transmissions to closer hops.
	Nonetheless, \textit{a priori} identifying energy efficient multi-hop routings is not trivial due to the unpredictable factors affecting the communication links in large LPWAN areas.
	In this paper, we propose \textit{epsilon multi-hop} (EMH), a simple reinforcement learning (RL) algorithm based on epsilon-greedy to enable reliable and low consumption LPWAN multi-hop topologies. Results from a real testbed show that multi-hop topologies based on EMH achieve significant energy savings with respect to the default single-hop approach, which are accentuated as the network operation progresses.
\end{abstract}


\section{Introduction} \label{Sec:Intro}

Low-power wide area networks (LPWANs) are wireless networks conceived for providing extensive communication ranges, reducing the energy consumption of end devices (STAs), and diminishing the operational cost with respect to traditional cellular networks. As a result, they are envisioned to be a key communication technology for a vast variety of Internet of Things (IoT) applications. 
LPWANs reach such a low power operation and extensive coverage range  by using the sub-1 GHz unlicensed, industrial, scientific and medical (ISM) frequency band, high processing gains, narrow bandwidths, and by sporadically transmitting packets at low data rates, which allows achieving very low sensitivities.

Most LPWAN solutions like LoRaWAN or SIGFOX are based on star topologies, where STAs directly transmit to the base station or gateway (GW), making them to heavily rely on transceiver's capabilities (e.g., available transmission powers, antenna gains or data rates). 
While this approach facilitates network designs and provides a robust centralized management, it usually leads to a shortening of the lifetime of STAs located far from the GW since they are most likely required to transmit using high power levels. In addition, the inclusion of STAs with limited transmission power is greatly compromised due to this range constraint. Moreover, long-range single-hop topologies lead to interference and packet collisions among uncoordinated devices, which importantly affects the reliability and scalability of networks with a large number of nodes~\cite{bankov2016limits, georgiou2017low}.

\begin{figure}[t]
	\centering
	\includegraphics[width=0.49\textwidth]{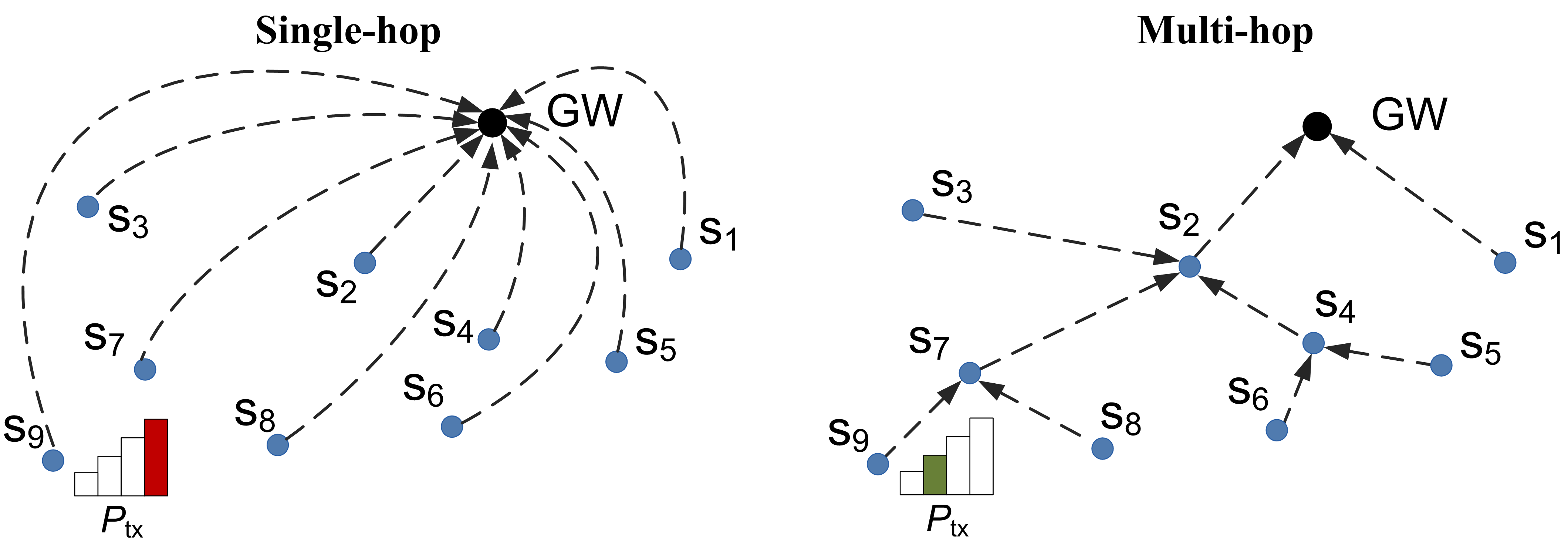}
	\caption{Single-hop vs. multi-hop topology. Note the power level reduction of STA $s_9$ when multi-hop is adopted.}    
	\label{fig:routing_example}
\end{figure}

Although multi-hop energy savings have shown its potential for wireless sensor networks \cite{kakitani2011comparing}, 
there is scarce literature on LPWANs on this topic~\cite{barrachina2017multi, barrachina2017learning}. These works reveal that multi-hop topologies can importantly extend LPWAN's lifetime by providing significant energy-savings to the STAs located farthest from the GW. In this regard, authors in~\cite{adame2018hare} present a novel LPWAN protocol stack enabling multi-hop communication in the uplink (HARE) that is able to achieve important energy savings in a real testbed.
One of the key challenges of multi-hop routing, however, is how to find both energy efficient and reliable links in a distributed way. Such difficulty results from the lack of global information required to make proper decisions. Instead, by exploiting the traditional centralized management of LPWANs, where a global view of the network is available at the GW (e.g., number of nodes, packet error rate, delay, etc.), the system is able to foresee if multi-hop routing strategies can outperform single-hop, and if so, reconfigure the network accordingly. Fig. \ref{fig:routing_example} shows the single-hop and a possible multi-hop topology on the same network deployment.

Indeed, it is difficult and fuzzy to determine \textit{a priori} the most energy efficient multi-hop routing for a given LPWAN. The main reason is that performance depends on too many factors such as the operation modes of the nodes (at microprocessor and radio modules) and the network deployment (e.g., location of the nodes, running applications, or environmental conditions). Hence, while deterministic rule-based solutions are not accurate for identifying energy efficient routings, machine learning approaches are appealing for such a task; especially when considering the scalability issues involved in huge LPWANs. Notwithstanding, a learning-based routing algorithm, if not properly set, could also entail significant extra consumption, and be consequently counterproductive since highly energy consuming topologies may be occasionally adopted~\cite{barrachina2017learning}. Therefore, the trade-off between the energy savings achieved with the most efficient routing and the cost of learning should not be overlooked.

As in any exploration/exploitation problem, reinforcement learning (RL) methods are appropriate due to their ability to cope online with such a tradeoff: \textit{i}) selecting the routing providing the best-known results (i.e, exploiting), or \textit{ii}) broadening the gathered knowledge about the performance of unexplored routings (i.e, exploring).	In this paper, we present results from a real testbed for assessing the energy savings achieved with \textit{epsilon multi-hop} (EMH), a Multi-Armed Bandits (MABs) $\epsilon$-greedy-based algorithm for learning energy efficient uplink routings in LPWANs. Namely, we assess the performance of EMH in an LPWAN operating under the HARE protocol stack, probing significant energy savings with respect to the single-hop approach.

%
%
%
%
%
%


\section{Overview of LPWANs} \label{Sec:overview}

\subsection{Communication challenges}

Most emergent LPWAN technologies only rely on the capabilities of low communication layers to achieve large single-hop coverage areas, disregarding multi-hop schemes already existing in other wireless networks. Star topologies are therefore predominant in LPWANs, where one central element (i.e., the GW) is the single responsible for configuring and managing the whole network. While simple and robust, this approach does not seem the most appropriate to face the following challenges~\cite{raza2017low}:

\begin{itemize}
	\item \textbf{Scalability and reliability}: since the propagation ranges are much higher, LPWANs cause interference at a much larger scale, creating bottlenecks in highly dense scenarios. Besides, most existing channel access mechanisms of LPWAN technologies resort to the use of ALOHA, which does not require much coordination between the AP and STAs~\cite{laya2016goodbye}. However, as the number of devices attempting to access the channel increases, so does the collision probability.
	\item \textbf{Flexibility}: current LPWANs are deployed, operated and managed in a completely uncoordinated manner, hindering new application purposes and/or possible network reconfigurations/upgrades.
	\item \textbf{Energy efficiency}: battery-powered LPWAN devices are currently lacking strategies beyond the PHY layer to extend their lifetimes, such as adaptive power control or advanced low duty cycle techniques combined with grouping strategies.
	\item \textbf{Quality of Service (QoS)}: since channel access is still randomized to some extent, no real guarantees in terms of QoS can be offered in LPWANs.
\end{itemize}

\subsection{Current LPWAN technologies}

While numerous LPWAN technologies have emerged in the last years, only some of them are able to combine long-range links and heterogeneous network topologies:

\begin{itemize}
	
	\item \textbf{HARE}, unlike other LPWAN technologies, is able to adopt uplink multi-hop communications without affecting data transmission reliability and achieving a notable energy consumption reduction~\cite{adame2018hare}. Multi-hop paths also involve intermediate STAs, which must be awake during the periodic association stages to execute the own distance-vector routing protocol.
	
	\item \textbf{LoRaBlink} incorporates multi-hop bi-directional communication enabling sensing and actuation \cite{bor2016lora}. Messages from nodes to the sink are directly flooded.

	\item \textbf{D7AP} networks consist of gateways and endpoints, and can optionally contain sub-controllers, thus also enabling tree topologies \cite{ayoub2018overview}. While gateways are permanently listening for packets, sub-controllers are allowed to sleep and are mainly used to relay packets. Lastly, endpoints can transmit asynchronously and wake up periodically to listen to possible incoming data. 
	\item \textbf{IEEE 802.11ah} includes in its specification a two-hop mode by using relays \cite{adame2014ieee}. Consequently, when transmitting to a closer relay instead to the AP, STAs reduce the transmission power level and use higher data rates, thus also shortening the transmission time, and consequently, the energy consumption.
\end{itemize}

\section{EMH: learning-based UL multi-hop routing} \label{Sec:learning}

In this section, we argue the need of learning proper routings for obtaining significant energy savings in real LPWANs and present the novel EMH approach.
To do so, let us define some terms regarding the UL routing for the sake of facilitating further explanation.

Any routing in a network can be represented by a vector $\vec{r}$ of size $n-1$, where $n$ is the number of nodes (GW and STAs) in the network, and whose element $\vec{r}(s)$ is the network address of the parent of STA $s$. The GW address is always set to 0 for simplicity.
In Fig. \ref{fig:routing_example} it is shown an UL multi-hop routing example with the corresponding routing vector $\vec{r} = (0,0,2,2,4,4,2,7,7)$. For instance, the parent of $s_5$ is $s_4$, i.e., $\vec{r}(5) = 4$.
The set $\mathcal{R} = \{\vec{r}\}$ is composed of all the possible UL routings that can be given in the network. Like in common LPWANs, we assume that every STA is capable (if required) to successfully communicate with the GW in a single-hop manner.

\subsection{The need of learning}

Due to the fact that the network performance depends on multiple factors such as the deployment of nodes, protocol stack, hardware, or environment conditions, it is hard and fuzzy to predefine proper UL routings beforehand. Therefore, even though experimentally tuning routing parameters can importantly enhance the energy savings in distributed approaches~\cite{adame2018hare}, in most of the cases its performance is sub-optimal. Moreover, such tuning approach comes at the cost of flexibility since the resulting configuration is deeply tied to the targeted scenario. 

In this regard, the problem of identifying the optimal routing can be modeled as a finite-horizon multi-armed bandit (MAB) problem due to its exploration/exploitation nature (i.e., the trade-off between exploring new knowledge or exploiting gathered knowledge) and the need of maximizing the lifetime of battery constrained STAs. While over-exploring routings prevents from maximizing the short-term reward in terms of energy savings, exploiting only partial knowledge prevents from identifying the optimal routing and maximizing the long-term reward accordingly. RL for wireless network has been previously covered in a number of papers like \cite{yau2012reinforcement,wilhelmi2017collaborative, wilhelmi2017implications}.

We propose EMH as a centralized learning-based routing approach that enables the GW to stochastically compute the routing table according to a MAB's $\epsilon$-greedy procedure. The goal of EMH is to minimize the energy consumption of the bottleneck STA (i.e., the STA that consumes the most) by exploring different UL routings. While simple to implement, this algorithm effectively serves to evaluate the impact of the different explored routings on the network's lifetime in a real-time manner.


\subsection{The EMH approach}

The well-known $\epsilon$-greedy method sets the randomness in action selection through a parameter $\epsilon$ that determines the probability of exploring a new action (already explored or not) rather than exploiting a previously explored one~\cite{watkins1989learning}. The simplicity of $\epsilon$-greedy together with the fact that no memorization of exploration specific data (e.g., counters or confidence bounds) is required~\cite{tokic2011value} are its main advantages with respect to other MAB methods. However, a substantial disadvantage of $\epsilon$-greedy is the complexity of determining the optimal initial value and the updating function of $\epsilon$.

In the EMH algorithm, a principal variation is included with respect to the regular $\epsilon$-greedy method: each action is explored just once.\footnote{Note that exploring one routing takes several energy consumption measures since they are averaged for improving the estimation accuracy.} Namely, since LPWAN deployments are characteristically static, we assume that the average energy consumed by the STAs in a certain UL routing does not significantly vary over time. Thus, we are able to set a deterministic (experimental) payoff to every routing by exploring it just once.

With respect to the testbed presented below in this work, the reward or payoff ($p$) provided by any possible action or routing in $\mathcal{R}$ may vary according to the channel condition (e.g., people crossing by or changing weather conditions). Therefore, in order to ensure enough accuracy of the payoff estimate ($\hat{p}$), we average the reward of each explored routing by measuring its corresponding energy consumption $K$ times. The pseudocode containing the main steps of EMH is depicted in Algorithm \ref{alg:egreedy}.

\begin{algorithm}[t]
	\footnotesize
	\SetAlgoLined
	\textbf{Input:} \\
	\text{  } $K$ \texttt{\textcolor{gray}{\#Number of averaging cycles}}\\
	\textbf{Initialize:} \\
	\text{  } $t \coloneqq 0$ \\
	\text{  } $\hat{p}(\vec{r})\coloneqq 0 \text{ for } \forall r \in \mathcal{R}$ \\
	\text{  } $\epsilon \coloneqq \epsilon_0$\\
	\While{active}{
		\texttt{\textcolor{gray}{\#New iteration}}\\
		$\vec{\gamma}_t \gets \texttt{estimate\_rssi}()$ {\texttt{\textcolor{gray}{\#RSSI from each STA}}} \\
		$\mathcal{A}_t \gets \{\vec{r} \in \mathcal{R} \mid \vec{\gamma}_t(s) \geq \vec{\gamma}_t(s') \text{ for } \forall (s,s')\}$ \texttt{\textcolor{gray}{\#Constraint}}\\
		$\mathcal{A}_t' \gets \{\vec{r} \in \mathcal{A} \mid \hat{p}(\vec{r}) = 0\}$ \texttt{\textcolor{gray}{\#Unexplored routings}}\\
		$\vec{r}_t$  $\begin{cases}
		\text{Explore: }\vec{r} \sim \mathcal{U}(\mathcal{A'}_t),              & \text{with prob. } \epsilon\\
		\text{Exploit: } \underset{\vec{r} \in (\mathcal{A}_t \setminus \mathcal{A}_t')}{\text{argmax }} \hat{p}(\vec{r}),& \text{otherwise}
		\end{cases}$\\
		$\bar{e}_\text{b}(\vec{r_t}) \gets \underset{s}{\max}\frac{1}{K} \sum_{k=1}^{K} e_{s,k}(\vec{r_t})$\\
		$\hat{p}(\vec{r_t}) \gets 1/\bar{e}_\text{b}(\vec{r_t})$\\
		$\epsilon \gets \epsilon_0 / \sqrt{t}$ \\
		$ t \gets t + 1$
	}
	\caption{Implementation of EMH in HARE. $\mathcal{U}(\mathcal{A}')$ is a distribution that randomly chooses any unexplored routing in $\mathcal{A}'$ uniformly at random.}
	\label{alg:egreedy}	
\end{algorithm}

\subsubsection{Estimating the single-hop RSSI}


the number of existing UL routings for networks of $n$ nodes is given by Cayley's formula.\footnote{In graph theory, Cayley's formula states that for every positive integer $n$, the number of trees on $n$ labeled vertices is $n^{(n-2)}$.} Specifically, $|\mathcal{R}| = n^{(n-2)}$. Hence, $|\mathcal{R}|$ grows extremely rapidly for large networks.	
In this regard, exploring routings without any predefined discrimination criteria could have a negative impact on the EMH performance. For instance, if considering a network deployment like the one shown in Fig.~\ref{fig:routing_example}, an alternative routing with $\vec{r}(1) = 9$ would be most likely sub-optimal since link $s_1$-$s_9$ probably suffers worst channel conditions than link $s_1$-GW. Consequently, $s_1$ will most likely suffer from higher energy consumption with respect to the original routing with $\vec{r}(1) = 0$.

In order to avoid exploring such \textit{naive} routings, we apply a received signal strength indication (RSSI) constraint stating that any children-parent link $(s,s')$ can only be performed if the RSSI received at the GW from the children is less or equal than the one received from the parent, i.e., $\vec{\gamma}(s) \leq \vec{\gamma}\text(s')$. Thus, we are able to significantly reduce the number of possible routings from $\mathcal{R}$ to $\mathcal{A} \subseteq \mathcal{R}$ by excluding those that do not comply with the \textit{RSSI constraint}. Note that we consider RSSI values rather than distances since channel conditions also depend on other deployment factors.

Accordingly, a preliminary step is conducted in each $\epsilon$-greedy iteration $t$. Basically, the GW estimates $\vec{\gamma}_t$ in the association phase, when STAs transmit directly to the GW (i.e., in a single-hop manner) asking for being associated to the network. With such a metric, the algorithm is able to discern what routings comply with the \textit{RSSI constraint} and identify the set $\mathcal{A}$.
EMH re-estimates $\vec{\gamma}_t$ in each iteration just in case the channel conditions have significantly changed since the network initialization. 

\subsubsection{Exploring or exploiting}

once $\vec{\gamma}_t$ is estimated, the algorithm decides whether to explore an unexplored routing or exploiting the best-known one according to the $\epsilon$ parameter. Specifically, with probability $(1-\epsilon)$ the algorithm picks the most energy efficient routing from the set of explored ones. That is, the already explored routing providing the highest estimated reward $\hat{p}$. Instead, with probability $\epsilon$, the algorithm picks uniformly at random an unexplored routing in $\mathcal{A}_t'$.

\subsubsection{Estimating the payoff}

after routing $\vec{r}_t$ is selected, the GW starts collecting the energy consumption measures of the STAs during $K$ HARE operation cycles. 
An important trade-off exists in this regard: the larger $K$, the more accurate the estimated payoffs, but the longer the time to identify the most energy efficient routing. The latter also entails the risk of exploring high consuming routings during larger periods of time.
After every cycle $k$, STA $s$ estimates the energy consumed during the cycle ($e_{s,k}$), generates a payload including the corresponding value, aggregates the payloads received from its children (if any), and transmits the packet/s with the payload/s to its parent. Once the $K$-cycles data collection phase finishes, the GW is able to determine the average energy consumed by every STA. Then, it sets the payoff estimate $\hat{p}$ corresponding to the current routing $\vec{r}_t$ as the inverse of the bottleneck node's average consumption.


\subsubsection{\protect\boldmath Updating the $\epsilon$ value}

once the payoff corresponding to the current routing $\vec{r}_t$ is estimated, the algorithm updates $\epsilon$. In our experiments, we use a time-dependent exploration rate $\epsilon_t = \epsilon_0 / \sqrt{t}$ with $\epsilon_0 = 1$ as suggested in~\cite{auer2002finite}. This $\epsilon$ setting entails substantially exploring in early stages and frequently exploiting afterwards, which is convenient for avoiding payoff local minimums. After updating the $\epsilon$ value, a new iteration begins with the single-hop RSSI estimation.

\section{Evaluation} \label{Sec:evaluation}

In this section, we evaluate the performance in terms of energy savings of the single-hop (SH) and EMH approaches. We first describe the testbed used for conducting the experiments along with the STA's energy consumption model. Then, we compare the performance of the aforementioned approaches. 

\subsection{Testbed}

\begin{figure*}[t]
	\centering
	\includegraphics[width=0.7\textwidth]{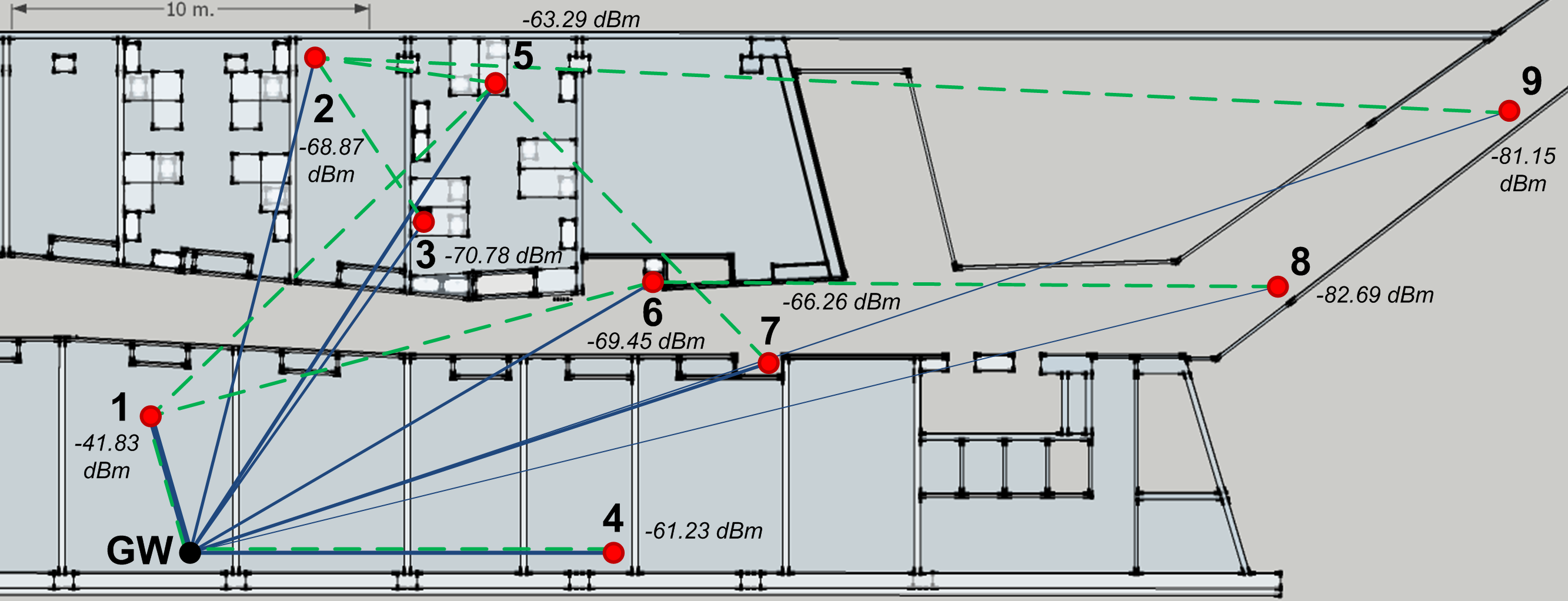}
	\caption{Testbed deployment. The weights of the SH edges correspond to the RSSI level perceived by the GW in the first association phase. The most energy efficient routing explored by EMH, $\vec{r} = (0,5,2,0,1,1,5,6,2)$, is drawn in dashed lines.}
	\label{fig:testbed_map}
\end{figure*}

\subsubsection{Deployment}

the performance evaluation of EMH and SH was performed in an indoor testbed located in one office building from Universitat Pompeu Fabra facilities. 9 Zolertia RE-Mote development boards/nodes~\cite{zolertiaRemote} acting as STAs were deployed throughout the offices and the main corridor, maintaining their location for all the experiments performed (see Fig. \ref{fig:testbed_map}).\footnote{More details such as the generated logs of the experiment are available at \url{https://github.com/sergiobarra/data_repos/tree/master/barrachina2018towards}.} Another Zolertia RE-Mote played the role of GW and was connected to a PC for logs generation. All devices ran Contiki 3.0 OS~\cite{dunkels2004contiki} as operating system and HARE as wireless communication protocol stack like the testbed from~\cite{adame2018hare}.\footnote{To the best of our knowledge, HARE is the only well tested LPWAN protocol stack specifically designed for UL multi-hop communications.} The selected radio duty cycle (RDC) sublayer was X-MAC, which defines sleeping periods for receivers and strobed preambles for transmitters~\cite{buettner2006x}.
The largest single-hop distance from the GW to STA \#9 was 45 meters. All operational tests were conducted considering no mobility. All STAs were powered by an 800 mAh battery except the GW, which was permanently powered by the PC.
STAs transmitted packets of 43 bytes every 2-minute cycle in a time division multiple access (TDMA) basis for group of contenders. In addition, the GW broadcast the routing at each iteration period so the STAs were able to identify their next-hop (or parent) and keep it until a new iteration was started.

Although LPWANs are characteristically composed of a large number of STAs located in outdoor scenarios, the presented testbed is sufficient to conduct a proof of concept providing significant results.\footnote{Note that larger LPWAN deployments have not been considered since they are expensive and hard to monitor. Nonetheless, the HARE protocol stack operation was already validated in outdoor environments~\cite{adame2018hare}.} In fact, since RSSI levels perceived by the GW are the main parameters used by the $\epsilon$-greedy approach, the actual position of the STAs and the channel conditions are always mapped to such parameters. That is, EMH is transparent to the actual LPWAN deployment.

\begin{table}[t]
	\footnotesize
	\centering
	\caption{Current values of the Zolertia RE-Mote platform at the different operational states.}
	\label{table:current_values}
	\resizebox{\columnwidth}{!}{
		\begin{tabular}{|l|c|c|}
			\hline                          & \textbf{Operational state} & \textbf{Current} \\ \hline
			\multirow{2}{*}{\begin{tabular}[c]{@{}l@{}}\textbf{Microprocessor} \\ ARM Cortex-M3\end{tabular}} & Processing (CPU) & $I_{\text{CPU}}=13$ mA\\ \cline{2-3} 
			&   Low power mode (LPM)       &    $I_{\text{LPM}}=0.4$ {\textmu}A   \\ \hline
			\multirow{3}{*}{\begin{tabular}[c]{@{}l@{}}\textbf{Radio Module} \\ TI CC1200 868 MHz \\ 2-GFSK, 50 kbps\end{tabular}}   &  Receiving (RX) & $I_{\text{RX}}=19$ mA      \\ \cline{2-3} 
			&   Transmitting (TX) & $I_{\text{TX}}=39-61$ mA      \\ \cline{2-3} 
			&   Sleeping (SL) & $I_{\text{SL}}=0.12$ {\textmu}A      \\ \hline
		\end{tabular}
	}
\end{table}

\subsubsection{Energy estimation}

the total energy ($e$) consumed by an STA is employed by two main elements: the microprocessor ($e_{\mu}$) and the radio power module ($e_\text{r}$). Specifically, $e_{\mu} = V_\text{DD} (t_\text{CPU} I_\text{CPU} + t_\text{LPM} I_\text{LPM})$ and $e_\text{r} = V_\text{DD} (t_\text{RX} I_\text{RX} + t_\text{TX} I_\text{TX} + t_\text{SL} I_\text{SL})$, where $V_{\text{DD}}$ is the supply voltage. The duration and current consumption corresponding to the operational states of the microprocessor and the radio module are $t$ and $I$, respectively. Table \ref{table:current_values} lists these states and the values of current consumption corresponding to the Zolertia RE-Mote. Notice that $I_{\text{TX}}$ value grows according to higher $P_\text{TX}$ transmission power levels (with a $P_\text{TX}$ operational range going from -16 to 14 dBm). We use the \texttt{energest()} function from Contiki to estimate the time an STA spends in each of the possible operational states for $K=10$ averaging measures.

\subsection{Results}


In order to assess the energy efficiency of SH and EMH, we use two main metrics: the cycle bottleneck energy in iteration $t$, i.e., $e_{\text{b}}(t)$, and the cumulated bottleneck energy until $t$, i.e., $\mathcal{E}(t)$. While the former refers to the energy consumed by the STA that has consumed the most in iteration $t$, the latter refers to the cumulated energy consumed by the STA that has historically consumed the most since iteration 1.
Therefore, the metric $e_{\text{b}}(t)$ serves to assess the performance in terms of energy efficiency of the routing being applied in iteration $t$ and assigning its corresponding reward. Instead, $\mathcal{E}(t)$ allows us to estimate the lifetime of the network since it is directly related to the remaining energy in the STA that has historically consumed the most. Note that, regardless of the considered packet transmission frequency (1 packet every 2 minutes cycle in this setup), a similar lifetime value in terms of iterations would be obtained since STAs consume very little when being in sleeping mode (\SI{0.12}{\micro A}). That is why we represent time in the x-axis of the plots in Fig. \ref{fig:experiment} in iteration units.

\begin{figure}[t]
	\centering
	\begin{subfigure}{0.48\textwidth}
		\centering
		\includegraphics[width=1\textwidth]{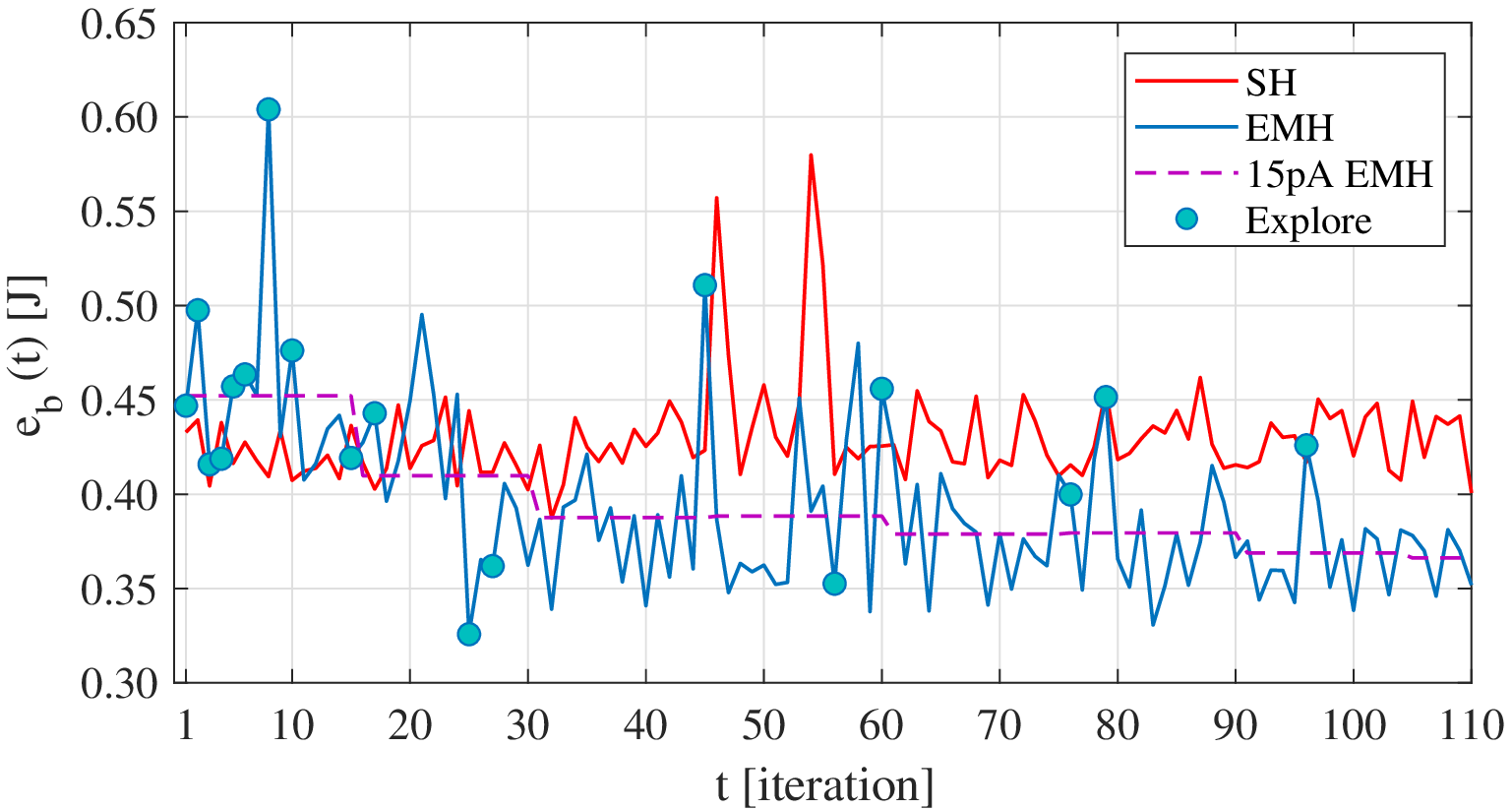}
		\caption{Average bottleneck energy for SH and MH. The curve 15pA-EMH refers to the average value corresponding to 15 consecutive measures of $e_\text{b}$.}    
		\label{fig:e_bottle}
		
	\end{subfigure}
	
	\vskip\baselineskip
	
	\begin{subfigure}{0.48\textwidth}
		\centering
		\includegraphics[width=1\textwidth]{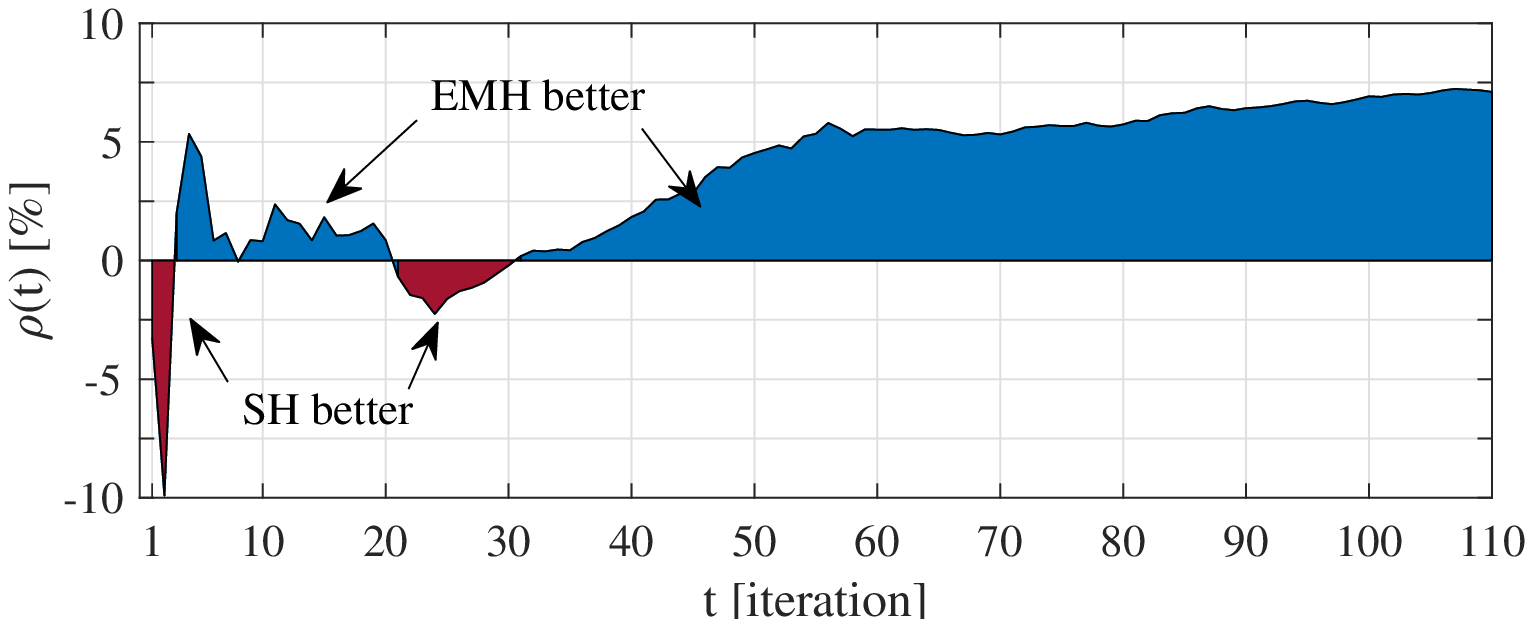}
		\caption{Saving ratio evolution. We highlight the periods when SH and EMH more energy-efficient in red and blue, respectively.}    
		\label{fig:saving_ratio}
	\end{subfigure}
	\caption[ ]
	{\small Performance of SH vs. EMH.}
	\label{fig:experiment}
\end{figure}

\subsubsection{Cycle bottleneck energy}

while in EMH unexplored routings are stochastically picked in exploration iterations according to probability $\epsilon$, in SH the same routing is used throughout all the experiment, i.e., $\vec{r}_t=\vec{0} , \forall t$. Accordingly, as shown in Fig. \ref{fig:e_bottle}, the cycle bottleneck tends to decrease as the experiment evolves when implementing EMH.

Regarding the payoff of each action, we can see significant variability for the same routing both in SH and EMH. The main cause is the dynamic nature of the indoor communication channel, which affects the routing performance. Specifically, once a communication link $(s,s')$ is assigned between a pair of nodes $s$ and $s'$, if channels conditions are not appropriate, several transmissions may be required to successfully deliver a packet due to noise and interference. In this regard, retransmissions entail important extra energy consumption in HARE networks due to the fact that both the transmitter and receiver must wake up again for retrying the communication. 
Such a phenomenon is more frequent in SH because of its inherent higher collision probability. Besides, in this type of routing, the bottleneck STA is normally located far from the GW because of its lower signal-to-interference-plus-noise ratio (SINR).

Nonetheless, by gathering $K$ energy measures per each routing, the GW is capable of estimating the corresponding payoff with sufficient accuracy to decide whether the routing is efficient or not. In fact, the tendency curve plotted in Fig. \ref{fig:e_bottle} shows the considerable energy reduction of the bottleneck STA with respect to SH, and its trend to slowly decay as more efficient routings are explored. We believe that in outdoor scenarios, since channel conditions are less dynamic, the $K$ value could be decreased while achieving better accuracy, thus outperforming the energy savings of the presented testbed.

In Fig. \ref{fig:testbed_map} it is shown the most energy efficient routing that the LPWAN explored during $T=110$ iterations. We note a clear multi-hop-like topology where packets are transmitted to intermediate STAs, which entails higher SINR values and corresponding reliability. Besides, the staggered wakeup pattern of HARE in multi-hop topologies allows reducing the channel contention and packet losses due to interference accordingly. 

However, since most of the energy consumed by the STAs is due to the operation in RX state, parent nodes tend to consume more energy as they need to wait for and decode packets from children. Hence, a balanced routing as the presented in this proof of concept is required. Besides, the dynamics and interrelations among STAs and the channel make it very difficult to determine beforehand whether a routing is energy efficient or not. In fact, in our experiments, we noticed that some routings following multi-hop approaches were clearly not energy efficient because, even though the average node consumption was low, one of the STAs consumed a lot compared to the rest. That is why learning is critical, especially for LPWANs with a huge number of STAs, where lots of different routings can be potentially established.

\subsubsection{Historic bottleneck energy}

the metric the metric that best maps the lifetime of the network is $\mathcal{E}(t)$ because of its direct relation with the remaining battery energy of the bottleneck STA. That is, any routing approach can be assessed in terms of energy saving by measuring its corresponding $\mathcal{E}(t)$, which is defined by
\begin{equation*}
\mathcal{E}(t) = \underset{s}{\max}\bigg(\sum_{t'=1}^{t} \frac{1}{K} \sum_{k=1}^{K} e_{s,k}(\vec{r}_{t'})\bigg)\text{.}
\end{equation*}

In order to compare the SH and EMH routing approaches, we show in Fig. \ref{fig:saving_ratio} the saving ratio between the energies consumed by their historic bottlenecks at iteration $t$, i.e.,
\begin{equation*}
\rho(t) = \frac{\mathcal{E}_{\text{SH}}(t) - \mathcal{E}_{\text{EMH}}(t)}{\mathcal{E}_{\text{SH}}(t)}\text{.} 
\end{equation*}

At the beginning of the experiments, due to the small amount of iterations performed and the frequent explorations, the routing heavily influences the historic bottleneck of EMH, and the saving ratio $\rho$ fluctuates accordingly. Instead, when the LPWAN is running for about 30 iterations, we note a more stationary behavior, where EMH clearly outperforms SH in terms of energy saving. Specifically, we achieve about 7 \% of saving in 110 iterations, which keeps growing over time.


\section{Conclusions} \label{Sec:conclusions}

Lowering energy consumption is critical for LPWANs due to their aim of supporting applications based on unattended, battery-powered devices. In this regard, multi-hop routings in the UL are starting to gain attention in the field. However, it is hazardous and sometimes counterproductive to predefine static routings prior to the deployment of an LPWAN.

In this paper we have proposed EMH, a centralized reinforcement learning (RL) algorithm for finding energy efficient routings in an exploration/exploitation approach. That is, while the network is normally operating, unexplored routings are stochastically chosen and assessed according to the bottleneck energy payoff function. Results from a HARE testbed with real LPWAN devices show that EMH achieves important energy savings with respect to single-hop topologies.

Finally, we envision that the use of centralized learning-based multi-hop routing will result in high energy savings in massive LPWANs (with up to thousands STAs) for two main reasons. On the one hand, multi-hop approaches are able to reduce the single-hop bottleneck energy consumed by those STAs located far from the GW, which are more likely to suffer from low SINR and other medium access issues like hidden and exposed node problems. On the other hand, with simple learning-based routing algorithms like EMH, we are able to find energy efficient routings that diminish the contention among STAs and build more reliable communication links. 


\bibliographystyle{unsrt}
\bibliography{bibliography}

\end{document}